%% file: sgs.tex
\documentclass[reqno]{article}
\usepackage{amsmath,amsfonts,amssymb,amsthm,xcolor,colortbl,tabularx,multirow,graphicx}
\usepackage{hyperref}
\usepackage[vertfit]{breakurl}
\usepackage[section]{placeins}
\usepackage[toc,page]{appendix}

\renewcommand\epsilon{\varepsilon}

\renewcommand\le{\leqslant}

\begin{document}

\title{Efficient subgraph-based sampling of Ising-type models with frustration}
\author{Alex Selby%
\thanks{alex.selby@cantab.net}
}
\date{25 August, 2014}
\maketitle

\newcommand{\SC}{\mathbf{S}}
\newcommand{\PP}{{\prime\prime}}

\begin{abstract}

Here is proposed a general subgraph-based method for efficiently sampling certain
graphical models, typically using subgraphs of a fixed treewidth, and also a related
method for finding minimum energy (ground) states. In the case of models with frustration,
such as the spin glass, evidence is presented that this method can be more efficient than
traditional single-site update methods.

\end{abstract}

\section{Introduction}\label{intro}

Given a graph $G$ with a weight $J_{ij}$ for each edge $(i,j)$, and variables, or
\emph{spins}, $\SC=\{S_i=\pm1|i\in V\}$ for each vertex, we can consider the energy
function defined by
\[H=H(\SC)=-\sum_{\text{edges}~(i,j)}J_{ij}S_i S_j.\]

Here the edges are considered to be undirected, so that if $(i,j)$ is in the sum, then
$(j,i)$ is not. We imagine that there is frustration in the model, i.e., many small
loops where the product of the weights $J_{ij}$ is negative. This typically arises when
$J_{ij}$ is itself chosen randomly, such as in the Edwards--Anderson spin glass.

This is the most straightforward case and so convenient for descriptions and
illustrations.  More generally, there could be more than two possible values, or
\emph{spin states}, for $S_i$ and there could be \emph{external fields} --- terms in
$H(\SC)$ depending on individual $S_i$.  In general, $H(\SC)$ can be the sum
of arbitrary functions of the form $J_{ij}(S_i,S_j)$ and $h_i(S_i)$ for the purposes of
the method described here.  The empirical runs described later will involve 16 states.

Two classical problems are considered here, the first of finding \textbf{ground states} or
minimum energy states, and the second of \textbf{simulating} or \textbf{sampling from} the
Gibbs/Boltzmann distribution.

Given an inverse temperature, $\beta$, the Gibbs/Boltzmann distribution over sets of spin
configurations is given as usual by
\begin{align*}
P(\SC)&=Z(\beta)^{-1}e^{-\beta H(\SC)},\qquad\text{with}\\
Z(\beta)&=\sum_\SC e^{-\beta H(\SC)}.
\end{align*}

Sampling from this distribution includes as a special case the problem of sampling the
ground states ($\beta=\infty$), and if there is frustration then finding a ground state is
known to be NP-hard for a general graph. Nevertheless, as seen in \cite{selbydwave1} and
\cite{selbydwave2}, ground states can be found quite efficiently for moderately large
graphs by searching over a covering set of low treewidth subgraphs.

The graph property \emph{treewidth} can be understood as the exponential complexity of
using dynamic programming to compute a locally-defined quantity. More precisely, this
statement applies to a graph $G$ and a property $\phi$ that is a collection of functions
$\phi_H$ for subgraphs $H\subset G$. Each $\phi_H$ is defined on the set of possible spin
states on $\partial H$, the set of vertices adjacent to $H$ but not in $H$. Suppose for
$H\subset H'\subset G$, $\phi_{H'}(\SC_{\partial H'})$ is equal to a simple combining
function on the collection $\phi_H(\SC_{\partial H})$, where $\SC_{\partial H}$ ranges
over the spin configurations on $\partial H$ that are compatible with the spin
configuration $\SC_{\partial H'}$. Suppose also that there are $s$ spin states for each
vertex, there are $m$ edges in $G$ and the treewidth of $G$ is $w$, then standard dynamic
programming using the tree decomposition will compute the property $\phi_G$ in $O(m\cdot
s^{w+1})$ steps. In what follows, we shall make use of two different choices of $\phi_H$:
the partition function $Z(\beta)$ for $H$, and a choice of random spins in $H$ according
to the Gibbs distribution.

There is a great deal of literature devoted to algorithms on families of graphs with
bounded treewidth, but here it is assumed that the desired graph of study does not have
bounded treewidth and the approach is to approximate it using bounded treewidth subgraphs.

The most common method of sampling Ising-type models generally involves Markov chain Monte
Carlo (MCMC), updating one spin at a time according to a random process dependent on its
immediate neighbours. More efficient methods using cluster updates, such as those of
Swendsen--Wang or Wolff \cite{Wolff}, are appropriate in models without frustration, but
do not work well in frustrated models. The method described here uses a different cluster
update method that works whether or not there is frustration.

A version of the subgraph sampling method was developed independently in 2013 by Decelle
and Krzakala \cite{decelle}, termed belief-propagation-guided Monte Carlo sampling. The
description given there apparently only considers the possibility of using trees for the
covering subgraphs (and provides a method for producing such trees at random). I believe
it is important to use a more general set of subgraphs, as described here, to be able to
sample (or find ground states) with large and difficult models. Evidence for this is given
in Section~\ref{comparison3}.

The sampling method described here is also similar to an earlier technique described by
Hamze and de Freitas \cite{hamze}, which focuses on a Markov Random Field with
observations attached. The method of \cite{hamze} only considers trees for subgraphs, and
only an exact partition of the graph by trees. As shown in Section~\ref{correct}, this
restriction is not necessary for the purposes of obtaining detailed balance, though under
these conditions the authors manage to prove rigorous bounds showing the efficacy of their
method.

The sampling method in \cite{fix} is similar to the one considered here, except that all
subgraphs of a given treewidth are allowed, not just induced subgraphs. I believe this is
significantly different from the method of induced subgraphs considered here and in
\cite{decelle}, for reasons given in Section~\ref{description}.

The primary method of comparison in this preprint will be wall-clock time, using a single
thread on a specific reference computer (an Intel Core i7-3930K CPU running at 3.20GHz),
except for the first comparison in Section~\ref{comparison1} where the method being
compared with was described by another party, and timings were not immediately
available. It may seem unusual to use something platform-dependent and
implementation-dependent like wall time rather than counting spin flips for example, but
not all spin flips are alike and I believe it is necessary to take seriously the facts (i)
that a spin flip isn't necessarily well-defined and (ii) that simpler methods are often
much more easily optimised. If a simple count of spin flips were used as a basis for
comparison, then this would tend to favour subgraph-based methods (at least as far as
sampling is concerned, if not ground state finding). This is because conventional methods
of flipping single spins can be executed efficiently without using any arithmetic
operations (which is how the comparison code used here works), whereas subgraph-based
sampling methods necessarily involve some kind of calculations to keep track of the
$Z$-values.

Each comparison here will involve combining subgraph-based sampling with parallel
tempering (also known as ``exchange Monte Carlo'', see, e.g.,~\cite{hukushima}) and
comparing this with single-site update methods also combined with parallel tempering. It
is well known that parallel tempering can significantly improve performance and we apply
it in both cases to ensure a useful comparison between the best available algorithms.

\section{Description of sampling method}\label{description}

Take a collection of induced subgraphs $T_1,\ldots, T_m$ of the graph $G$. Recall that an
induced subgraph is the restriction of $G$ to a particular subset of vertices, and
contains all the edges of $G$ between those vertices. In other words, if both of the
endpoints of an edge of $G$ are in some $T_i$ then the edge itself must be in $T_i$.  We
also require that $\cup T_i=G$, i.e., every vertex and edge is represented in some $T_i$.

We consider induced subgraphs only, as this ensures that when we sweep over a subgraph the
spins of all its neighbouring vertices can be held fixed.  It is possible to ignore this
restriction and still use the method for a non-induced subgraph, fixing the spin value
from the pre-updated graph to use as the neighbouring value during the subgraph
update. However, then monotonicity would be lost: at $\beta=\infty$ an induced subgraph
update is guaranteed not to increase the global energy, but this is no longer true for
non-induced subgraphs. I believe this will lead to poorer performance for non-induced
subgraphs as the $\beta=\infty$ point is the most difficult and the difficulty of
obtaining a ground state is a guide to the difficulty of obtaining low temperature / high
$\beta$ samples. The approach of \cite{fix} is to choose the best maximal subgraph of a
given treewidth (not necessarily induced, so at treewidth 1 this is a spanning tree),
trying to minimise the weight of the edges not in the subgraph, thereby making the
approximation as good as possible.

The idea is that $T_i$ should be chosen to be easy to solve exactly (in the sense of
finding its ground states, or calculating its conditional Gibbs distribution). A good
choice would be to only use subgraphs of a given treewidth. (It is not necessary to take
all subgraphs of a given treewidth.) For simplicity, the examples here are all illustrated
by the case of treewidth 1, i.e., $T_i$ will all be trees, but it should be remembered
that $T_i$ will have some non-trivial treewidth $w$ in general.

Given an induced subgraph, $T$, and a spin configuration, $\SC_{G\setminus T}=\{S_i|i\in
G\setminus T\}$ defined on the remainder of $G$, we can condition $P()$ on
$\SC_{G\setminus T}$ to get a distribution, $P_T(\cdot|\SC_{G\setminus T})$ over the spin
configurations, $\SC_T=\{S_i|i\in T\}$, over $T$.

The Monte Carlo step is to take $T$ to be a random $T_i$ from our fixed collection and
then replace $\SC_T$ with a random configuration chosen according to the conditional
distribution $P_T(\cdot|\SC_{G\setminus T})$. This defines a sampling procedure, subject to the
usual Monte Carlo caveats regarding burn-in and waiting for independent samples.

\vskip5pt
We need to show

\begin{itemize}
\item that this operation has the correct invariant distribution,
\item that this operation is efficient to perform, and
\item that the performance in terms of equilibration time can be better than traditional methods.
\end{itemize}

We'll take these in turn. The first two items are simple to demonstrate, but the last item
is more open-ended and will be evaluated in different ways in separate sections of this
preprint.

\subsection{Correctness of invariant distribution}\label{correct}

To show detailed balance, consider the flux from configuration of spins $\SC$ to
$\SC'$. Let us write $\SC\Delta\SC'$ to mean $\{i\in V|S_i\neq S'_i\}$. Then the flux
$\SC\to\SC'$ is equal to
\[P(\SC)\frac{1}{m}\sum_{T|\SC\Delta\SC'\subset T}P_T(\SC'_T|\SC_{G\setminus T}),\]
 where the sum is over subgraphs in our fixed collection that include $\SC\Delta\SC'$. This is equal to
\[P(\SC)\frac{1}{m}\sum_{T|\SC\Delta\SC'\subset T}\frac{P(\SC')}{\sum_{\SC''|\SC''_{G\setminus T}=\SC_{G\setminus T}}P(\SC'')}.\]
But $\SC\Delta\SC'\subset T$ is the same as saying $\SC_{G\setminus T}=\SC'_{G\setminus T}$, so the condition on $\SC^\PP$ in the sum in the denominator, $\SC''_{G\setminus T}=\SC_{G\setminus T}$, is equivalent to $\SC^\PP_{G\setminus T}=\SC'_{G\setminus T}$, and the whole expression is thus symmetric under interchanging $\SC$ and $\SC'$, proving detailed balance.

\subsection{Efficiency of calculation}\label{efficient}

For clarity we'll scrutinise the case where the $T_i$ all have treewidth 1, that is they
are trees. In general, if $T_i$ had treewidth $w$ then the time taken to draw a random
instance from $P_T(\cdot|\SC_{G\setminus T})$, or to calculate
$\sum_{\SC^\PP|\SC^\PP_{G\setminus T}=\SC_{G\setminus T}}P(\SC^\PP)$, would be roughly
$s^{w+1}|\mathcal{E}(T,G)|$ elementary operations, where $\mathcal{E}(T,G)$ is the set of
edges with at least one end in $T$ and $s$ is the number of (spin) states of each
vertex. For illustration we show how it works with $s=2$, $w=1$, though results reported
in later sections use up to $s=16$, $w=2$.

We proceed inductively over the vertices of $T$ from its leaves inwards in the usual
manner of dynamic programming based on a tree decomposition.  After $r$ steps, we have a
collection, $U_1,\ldots,U_m$ of connected subtrees of $T$ of total size $\sum_i|U_i|=r$,
with $T\setminus\cup_i U_i$ being connected. For example:

\begin{figure}[h]
\includegraphics{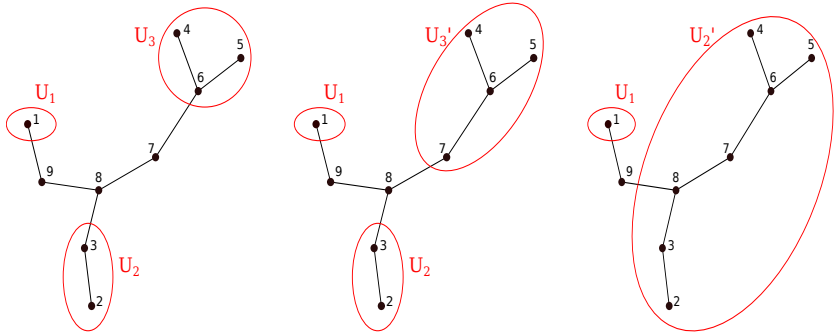}
\caption{Three steps in the inductive evaluation and sample-choosing of a subgraph}
\label{treedecomp}
\end{figure}

The above vertex numbering shows one of many possible orders in which the vertices may be
processed, and three successive steps are illustrated. The rule is a vertex may be
processed only when at most one of its neighbours is unprocessed. Note that the rest of
the graph $G\setminus T$ is suppressed in the above picture, but must be taken into
account as a ``background field" in the inductive calculation as it contributes to the
conditional probability of the tree by interactions along edges between $T$ and
$G\setminus T$.

There are two quantities that need to be maintained inductively, for each value $S_b$ of
the boundary spins of $U_i$ (the boundary of $U_i$ being vertices adjacent to $U_i$ that
are not themselves in $U_i$; in this tree case there is only one such vertex):
\begin{itemize}
\item The total $Z$ value for $U_i$ given $S_b$ and $\SC_{G\setminus T}$.
\item A choice of spins of the vertices in $U_i$, representing a sample of the distribution $P()$ conditioned on $S_b$ and $\SC_{G\setminus T}$.
\end{itemize}

These two quantities are easy to maintain inductively when a vertex is added to a $U_i$. To illustrate the process, the example of the above pictures is traced in some detail. For $U_3$ in the first picture there will be, for each of the two possible values, $\pm1$, of $S_7$
\begin{itemize}
\item the value $Z_{U_3}$ defined as:
\[Z_{U_3}(S_7)=\sum_{S_4,S_5,S_6}e^{-\beta H_{U_3}(S_4,S_5,S_6, S_7, \SC_{G\setminus T})}\]
where $H_{U_3}$ contains the contribution to the energy from edges meeting the vertices
$4$, $5$ and $6$ of $U_3$. These edges will join vertices $4,5,6$ to each other, to vertex
$7$, and to $G\setminus T$, but they can't meet the other vertices of $T$ by
construction. It should be remembered that the above expression for $Z_{U_3}$ is just a
definition, not the method by which it was calculated.

\item A choice of spins $\mathcal{C}_{U_3}(S_7)=(S_4,S_5,S_6)$
\end{itemize}

Moving from the first of the above pictures to the second, vertex $7$ is incorporated into
$U_3$ and $8$ is the new boundary vertex. It is seen that $\Delta_H=H_{U'_3}-H_{U_3}$
only depends on $S_7$ and $S_8$ (and spins over $G\setminus T$, which we suppress), and
no other vertices of $T$. So
\begin{itemize}
\item the new value $Z_{U'_3}$ is easily calculated as
\[Z_{U'_3}(S_8)=\sum_{S_7=\pm1}e^{-\beta\Delta_H(S_7,S_8)}Z_{U_3}(S_7),\]
\item and the new choice of spins is given by
\[
\mathcal{C}_{U_3'}(S_8)=\begin{cases} S_7=-1, (S_4,S_5,S_6)=\mathcal{C}_{U_3}(-1)&\text{with prob.}\;Z_{U'_3}(S_8)^{-1}Z_-\\
                                      S_7=+1, (S_4,S_5,S_6)=\mathcal{C}_{U_3}(+1)&\text{with prob.}\;Z_{U'_3}(S_8)^{-1}Z_+\\
\end{cases}
\]
where $Z_{\pm}$ are the summands in the above expression for $Z_{U'_3}(S_8)$.
\end{itemize}

When the new vertex involves fusing more than one $U_i$, the expression for the new $Z$
will involve products of the old $Z$s. In the example above, in the transition from the
second to third picture, $Z_{U'_2}$ is built up of the sum of terms of the form
$e^{-\beta\Delta_H}Z_{U_2}Z_{U'_3}$.

In practice this algorithm can be easily and compactly implemented, with the spin choice
being represented by a pointer. This is a small difference from the approach of
\cite{decelle} where a second reverse pass is used to construct the sample values for the
spins of the subgraph (which will be a tree).  Some care is needed for a fast
implementation since straightforward methods of dealing with the inevitable large
numerical range of values involve computing logarithms and exponentials and generating
random numbers in the inner loop, or the loop one level outside it, which would be rather
slow. These can be avoided, for example by using look-up tables for the exponentials,
rescaling the $Z$-values when necessary, and storing and judiciously reusing random
numbers. See Appendix~\ref{AppendixB} for further discussion of numerical aspects. The
code used to produce the results below is available from \cite{mygithub}.

\section{Chimera graphs}\label{chimera}

Experiments in this preprint were carried out on Chimera graphs (\cite{dwavechimera},
\cite{boixo1}) of various sizes. These were chosen because originally the aim was to
compare (\cite{selbydwave1}, \cite{selbydwave3}) classical optimisation with that of
D-Wave hardware \cite{boixo1} --- a quantum device whose current implementation is based
on a Chimera graph. This preprint will not contain any comparisons aginst D-Wave hardware
as its aim is look at classical optimisation and simulation techniques in their own right.

\begin{figure}[h]
\includegraphics{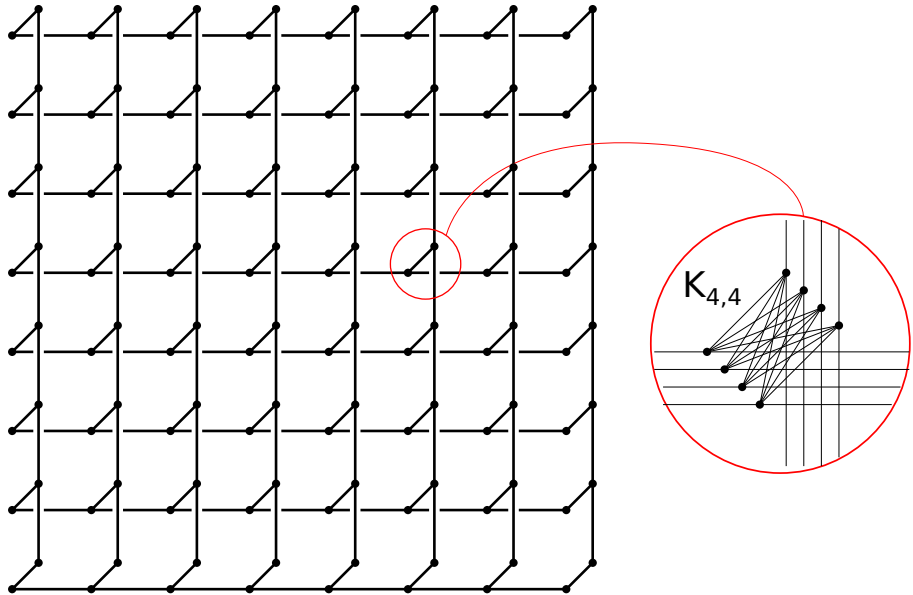}
\caption{$8\times8$ Chimera graph shown in collapsed view with 128 ``big vertices'', the inset showing a portion of the full 512-vertex graph}
\label{chimeraexample}
\end{figure}

An $n\times n$ Chimera graph, $C_n$, consists of $8n^2$ vertices arranged as $n^2$
complete bipartite graphs $K_{4,4}$. We shall use the notation $N=8n^2$
throughout. Writing the bipartite decomposition of $K_{4,4}$ as $A\cup B$, the $n^2$
graphs of the form $A$ are pointwise connected horizontally in rows, and the graphs of the
form $B$ connected vertically. An $8\times8$ Chimera graph is illustrated in
Fig.~\ref{chimeraexample}. The separate 4-vertex $A$ and $B$ graphs can be thought of as
\textbf{``big vertices''} (a term used throughout this preprint) in a simpler collapsed
graph with $2n^2$ vertices.

The Chimera graph is highly non-planar in that its genus is at least $(4/3)n^2+O(n)$,
which follows from the fact that the complete graph $K_{4n+1}$ can be minor-embedded into
$C_n$. This means that matching techniques \cite{planar} that enable ground states of
planar, and by extension low genus, graphs to be found in polynomial time are not
applicable here.  On the other hand, while the Chimera graph is not simply two
dimensional, it isn't fully three dimensional either. In \cite{katzgraber} it is shown
that in a certain scaling limit, $C_n$ behaves like a planar graph in that it has no
positive critical temperature.  This result is apparently at odds with the fact that
$K_{4n+1}$ is minor-embeddable in $C_n$ and should have a positive critical temperature
because a complete graph is effectively infinite dimensional.  However, these facts can be
reconciled because $|K_{4n+1}|/|C_n|\sim\sqrt{2/N}$ and \cite{katzgraber} is not concerned
with features on the scale of $1/\sqrt{N}$. Furthermore \cite{katzgraber} considers a
uniform weight distribution on the edges of $C_n$ which would not translate to a uniform
weight distribution on the edges of $K_{4n+1}$.

The treewidth of $C_n$ is $4n$, or $n$ in terms of big vertices. In practice, $C_n$ can be
exhaustively searched using a simple treewidth-based method to find either ground states
or perfect Gibbs samples at inverse temperature $\beta$, for $n$ up to 8 on a normal
desktop computer. Larger sizes rapidly present problems in terms of memory as well as time
for this method.  It is possible that branch-and-cut methods could be combined with
heuristic methods to increase beyond $n=8$ the largest $C_n$ that can be exhaustively
searched, but that is not explored here.

Following \cite{ronnow}, the class of random instances \textbf{Range $\mathbf{r}$} is defined by
choosing each edge coupling randomly from the set $\{-r,-r+1,\ldots,-1,1,\ldots,r\}$. The
instances considered here are Range 1 in Section~\ref{comparison1} and Range 7 in other
sections.

\section{Comparison - equilibration time and Binder ratio}\label{comparison1}

In \cite{katzgraber} the authors study, inter alia, the critical behaviour of bimodal
($J_{ij}=\pm1$, same as Range 1) spin-glass on the Chimera graph. They make a prediction
for its universality class, and test this by showing that the expectation of the Binder
ratio as a function of suitably-scaled temperature is independent of the size $N$ of the
underlying graph (figure 2, upper, of \cite{katzgraber}).

This gives us an opportunity to compare the subgraph-based sampling methods described here
with the more standard, though well-tuned, Monte Carlo methods used in the paper.

There are now two levels of probability space: the space of random $J_{ij}$ (known as the
``disorder"), and for a given choice of $J_{ij}$, the space of spins, $\SC$ (averaging
over which can be referred to as taking a ``thermal average").

In \cite{katzgraber}, parallel tempering is used, which greatly improves convergence in
this sort of problem. It is a kind of Monte Carlo meta-method and can be used ``on top" of
another Monte Carlo method. \cite{katzgraber} uses a standard Monte Carlo sweep as the
subroutine. For comparison we also use parallel tempering, but instead base it on top of
subgraph-based sampling.

The comparison we examine is with $p=0.5$, $N=512$ in the notation of
\cite{katzgraber}. That is, the graph is the Chimera graph of order 8 and $J_{ij}$ are IID
$\pm1$. (For avoidance of doubt, it is assumed here that the Hamiltonian given there in
formula (1) as $-\sum_{i,j=1}^N J_{ij}S_iS_j$ was intended as $-\sum_{i<j}J_{ij}S_iS_j$,
otherwise the undirected edge weights $J_{ij}+J_{ji}$ would effectively be taken from the
set $\{-1, 0, 1\}$, not $\{-1, 1\}$ as they are meant to be.)

The choice of temperatures used here covers a similar range (0.2 - 2) to that specified in
table 1 of \cite{katzgraber} (0.212 - 1.632). The temperature choice here was decided upon
by trying to make the exchange probabilities between adjacent temperatures in the parallel
tempering method all equal to 0.6, which ended up requiring 25 temperatures for the range
0.2 - 2. (As it turned out, these probabilities got a little higher than that for the
bottom two temperatures. In retrospect, this exchange probability of 0.6 may have been a
bit higher than optimal.)

For a given disorder, \cite{katzgraber} takes two independent random spin configurations,
$\SC$ and $\SC'$ and defines the spin overlap $q=(1/N)\sum_iS_iS'_i$. Then the quantity of
interest is its excess kurtosis, the Binder ratio

\[g_q=\frac{1}{2}\left(3-\frac{\langle q^4\rangle}{\langle q^2\rangle^2}\right)\]

Here $\langle.\rangle$ denotes thermal average and disorder average. The interest is in
the thermal average, as it is trivial to sample $J_{ij}$ to obtain the disorder average.

At very low temperatures, assuming the ground state is close to unique, we might expect
$q$ to take the values close to $+1$ or $-1$ according to whether $\SC'$ happened to hit
the same ground state as $\SC$ or its negation. This would make $g_q$ close to 1. At high
temperatures, $S_i$ will be independently $\pm1$, which makes
$q\approx2(1/N)B(N,1/2)-1\approx N(0,1/N)$ and so $g_q\approx0$. This is what we see, with
$g_q$ apparently decreasing smoothly at intermediate temperatures indicating the lack of a
phase transition at $T>0$, at least in the scaling limit as $N\to\infty$.

The subgraph-based method used here in this experiement was the Gibbs analogue of the
method known as strategy 3 in Appendix B of \cite{selbydwave1} and as PT-TW1 in Sections
\ref{comparison2} and \ref{comparison3}. It uses the collapsed Chimera graph of 128
aggregate vertices with 16 spin states each, and there is a prescribed set of 32 trees.

The relevant question in this simulation is how many steps the Monte Carlo process takes
to equilibrate, i.e., for the associated Markov chain to reach something close to an
equilibrium distribution. For each disorder, a pair of spin states is initialised
uniformly at random at each temperature. Then $R$ exchange parallel tempering steps are
performed, which involves $R$ tree-steps for each temperature. At that point the states
are deemed to be in thermal equilibrium and they are run for $R$ further Monte Carlo steps
during which the thermal averages $\langle q^2\rangle$ and $\langle q^4\rangle$ are
evaluated.

This whole process (including disorder average) was repeated for increasing values of $R$,
($250, 500, 1000, 2000, \ldots$) until the final values of $g_q$ appeared to stabilise
within acceptable error margins (up to 0.01 absolute). It turned out this required
$R=1000$, i.e., 1000 tree-steps. As far as I can gather, in terms of elementary
operations, such a tree-step should be approximately comparable to about 100 standard
Monte Carlo sweeps, when both are optimised, making about 100,000 sweep-equivalents for
equilibration. This compares with the $2^{21}$, or about 2,000,000 sweeps required in
\cite{katzgraber}, so it appears there might be a significant advantage with this
approach; but see also the discussion below.

\begin{figure}[h]
\input{Comparison-Katzgraber.tex}
\caption{Log Binder ratio vs rescaled temperature at N=512 after 5964 anneals --- comparison with \cite{katzgraber}.}
\label{brcomp1}
\end{figure}
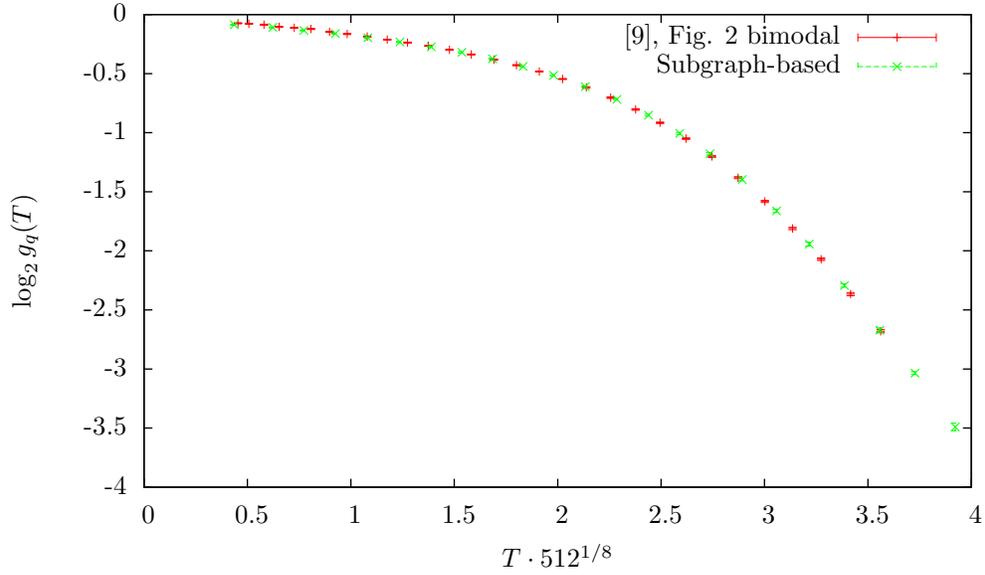

\begin{figure}[h]
\input{Comparison-Katzgraber-errorsx10.tex}
\caption{Log Binder ratio vs rescaled temperature at N=512 after 5964 anneals --- errors multiplied by 10.}
\label{brcomp2}
\end{figure}
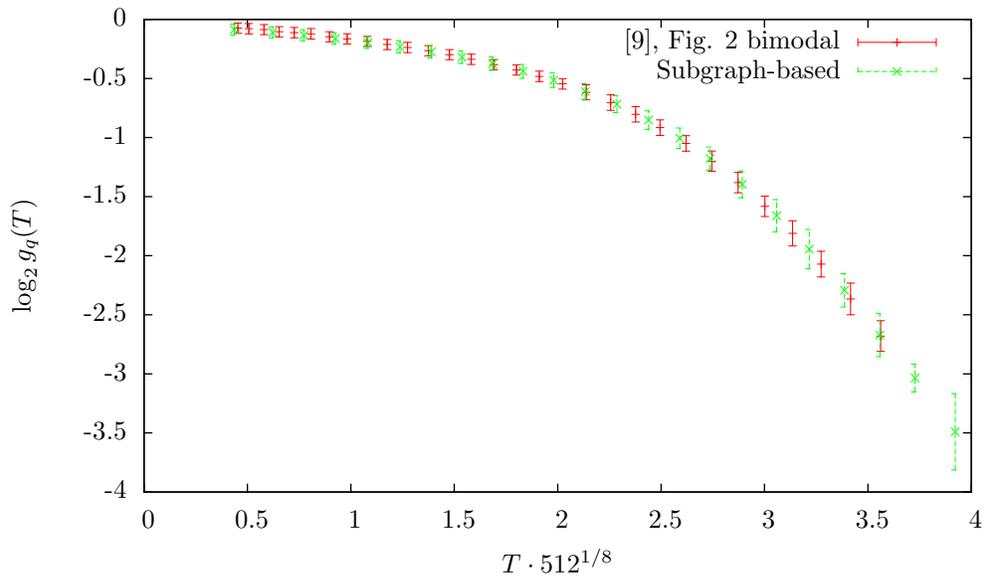

The graph in Fig.~\ref{brcomp1} shows the results of the method described here applied to
the bimodal (Range 1) example problem of~\cite{katzgraber} with $N=512$. The graph in
Fig.~\ref{brcomp2} shows the errors artificially amplified by a factor of 10, since they
are too small to be seen clearly at their true scale. As can be seen, there is good
agreement with the results given in Fig.~2 of ~\cite{katzgraber}, with similar
uncertainties. In both cases the results are averages over 5964 disorders. These graphs
serve as a check that the method described here is functioning correctly. They do not
compare performance, because that is hidden in the number of equilibration steps required
for each disorder.

Returning to the performance comparison, as alluded to above, there are some problems with
this comparison. In particular, we are making the following assumptions:
\begin{itemize}
\item that there is an equivalence between sweeps and tree-steps
\item that the accuracy (used as a termination criterion) is comparable
\item that the parallel tempering algorithms are tuned in the same way, or if not,
that the impact of any differences is negligible
\end{itemize}

The next sections describe how we have attempted to construct new comparisons that compare
the two approaches on as fair a basis as possible.
 
\section{Comparison of equilibration time using ground state values as the termination criterion}\label{comparison2}

To make a much more careful and controlled comparison of subgraph-update-based sampling
with site-update-based sampling, we choose a particular problem, and try to simulate it
using versions of these two methods that are as nearly identical as possible. As a matter of
notation:

\begin{itemize}
\item \textbf{SGS} shall mean subgraph-update-based sampling in general, as described in
  Section~\ref{description}. In the example below, we shall consider \textbf{PT-TW1},
  parallel tempering using treewidth 1 in the ``big vertex'' graph.
\item \textbf{SSS} shall mean single-site-update-based sampling. In other words, the
  traditional method whereby each spin variable is updated depending only on its immediate
  neighbours. (To be clear, this category includes multispin methods, such as that
  described in \cite{isakov}. Even though these methods operate on more than one spin at a
  time, the outcome is the same as operating on spins individually.)  The comparison below
  will use \textbf{PT-TW0}, parallel tempering using a conventional update based on
  immediate neighbours, in terms of big vertices.
\end{itemize}

The particular problem set chosen is that of the Chimera graphs of sizes $6\times6$,
$8\times8$, $10\times10$ and $12\times12$ (number of spins 288, 512, 800 and 1152
respectively). For these graphs, the couplings on the edges are chosen uniformly from the
14 possibilities $\pm1, \pm2, \ldots \pm7$, there are no external fields and each spin can
be $+1$ or $-1$. This mimics the ``Range 7" (harder) example set from \cite{ronnow}.

In fact the $J_{ij}$ used were half the values stated above, i.e., chosen from
$\pm\frac12, \pm\frac22, \ldots, \pm\frac72$, so that the energy quantum, the smallest
possible change in energy due to a change in $S_i$ for a given disorder $J_{ij}$, is $1$
rather than $2$. Of course this scaling factor doesn't fundamentally change anything
because $\beta$ can always be rescaled, but we state it explicitly to allow the reader to
interpret the numbers in what follows, where we mention specific values of $\beta$ and
maximum allowable absolute errors in energy.

To make an interesting and fair comparison we would ideally compare the best SGS-based
method against the best SSS-based method, in so far as that makes sense. Of course, we
don't necessarily know the best methods, but parallel tempering currently stands out as
one of the best methods known for equilibration of such frustrated models.

\subsection{Parallel Tempering parameters}\label{ptparams}

For each problem size, ($6\times6$, $8\times8$, $10\times10$ and $12\times12$), we choose
100 random disorders, and for each of these we determine the time required for the SGS-
and SSS-based methods to equilibrate to a reasonable accuracy. The principal statistic
comparing SGS with SSS is simply the respective totals of these times, though other
comparisons may be considered.

It may be argued that fixing the required accuracy for each disorder does not match the
likely underlying objective, which is to get a good expectation over disorders. It may be,
for example, that it is more efficient to spend less time trying to equilibrate the most
difficult disorders, allowing them to be less accurate, on the grounds that they will be
compensated for by the majority of more accurate disorders. I do not believe this kind of
optimisation technique would help a great deal for two reasons. First, by the nature of
the exponential decay of the subleading eigenvalue of the Monte Carlo iteration, there
should be a fairly sharp change from an inaccurate result to an accurate one as the number
of equilibration steps crosses the characteristic threshold for that particular
disorder. That means that one can't afford to use much less than the proper number of
equilibration steps, otherwise the results would be very inaccurate and swamp the accurate
results from other disorders.  Second, scrutinising the results here, though there are
certain disorders that are considerably harder than the others, these still don't
represent an overwhelming proportion of the total equilibration steps expended over all
100 disorders.

The set of temperatures is determined by fixing an effective absolute zero at
$\beta=\beta^*=20$ and aiming for a constant transition acceptance rate, $0.25$, between
neighbouring $\beta$s. The value $\beta=20$ is sufficiently large (for the class of
disorders considered here) that there is only a negligible chance of a state at this
inverse temperature not being the ground state. \cite{katzgraberemc} describes such a
constant acceptance rate as in general ``not too bad, but not optimal". The temperature
set here is determined in advance of the actual simulation using a method based on the
average energy and heat capacity at a range of different temperatures. The acceptance
rates during actual simulations match the target value reasonably well, almost always
lying between $0.2$ and $0.3$.

Having fixed the maximum $\beta$, and having fixed the spacing between the $\beta$s by
deciding on the transition acceptance rate, the remaining parameter to be decided is the
minimum $\beta$, or equivalently the number of temperatures. This is determined by trying
a range of different possible values on a trial set of disorders and seeing which number
of temperatures requires the fewest steps on average to equilibrate. It is found that SGS
requires slightly fewer temperatures than SSS for a given problem size, thus SSS will end
up with some ``bonus information" about high temperature states. However, it is not given
credit for this, and all that is compared is the time required to estimate the ground
state energy (and so presumably low temperature states) to a given accuracy. The
justification for this is that it is assumed that the main interest lies in colder
temperatures (strong coupling, high $\beta$), since higher temperatures are relatively
easy to simulate using any sensible method. Full details of the sets of temperatures used
are in Appendix~\ref{AppendixA}.

\subsection{Determination of Equilibration}

Equilibration is determined using the following method. Starting from a set of uniformly
random states at each temperature, $n$ Monte Carlo steps are performed. Each such step
consists of doing a single-temperature step at each temperature and then attempting to
perform exchanges. After that, $n$ more steps are performed during which any observables
can be measured and the energy at $\beta^*$ is averaged to make a sample value $E_1$. This
whole process is repeated $25$ times starting each time from random initial states, and
the average value $E=(E_1+\ldots+E_{25})/25$ is formed. If this is within a chosen
threshold (taken to be $0.2$ here) of the smallest energy observed at any point so far,
$E_\text{min}$, then it is deemed that $n$ steps are sufficient for equilibration. (It is
possible to simultaneously test, for each $m$, whether the smaller number of steps $m$
would have sufficed in not much more time than it takes just to test $n$ itself.)

This procedure relies on a number of assumptions. First, that $E_\text{min}$ is the true
ground state energy. Empirical evidence strongly suggests that it is, but for the purposes
of comparison it may not matter too much if it isn't, provided that the same value of
$E_\text{min}$ is used for both SSS and SGS, and this can be checked. Second that
$\beta^*$ is the hardest value of $\beta$ to equilibrate and the states at other
temperatures will have equilibrated if the state at $\beta^*$ has. Even if this turns out
not to be an accurate assumption, then at least SSS and SGS are still being compared on an
equal basis. In any case, it is assumed that the lower temperatures are the objects of
interest and the higher temperatures are a means to this end. The third assumption is that
the number of restarts ($25$) is sufficiently large to get a good estimate of the required
number of equilibration steps. The number $25$ was chosen to limit the total computer time
for all experiments to a week or so, but in fact it is on the small side and there is a
noticeable variance in the estimate for the required number of equilibration
steps. However, when the estimate is averaged over 100 disorders, this variance becomes
tolerably small.

\subsection{Timing}

The aim is to compare wall time, though for practical reasons we break this down into the
product of the number of parallel tempering (PT) steps and the time per PT-step. A PT-step
includes a sweep at each temperature and the attempting to do an exchange for each pair of
neighbouring temperatures. The time for such a step is liable to change if low-level
spin-flip algorithms are optimised, if the computing device changes, or if there are other
processes running on the computer. Separating out the wall time into a
platform-independent step count and a simple low-level timing frees us up to do the
step-counting runs in any environment with code at different stages of optimisation, means
that we only have to measure the timings once under controlled conditions, and enables us
to consider the effects of further optimisations. It is hoped that the ratio
$t_\text{PT-TW1}/t_\text{PT-TW0}$ of times per step for the respective methods is fairly
robust and won't vary too much across different platforms, though it is a significant
assumption that this ratio is stable under further optimisation of the respective
low-level spin-flip algorithms. The times $t_\text{TW1}$ and $t_\text{TW0}$ were measured
on a reference computer (an Intel Core i7-3930K CPU @ 3.20GHz) where both spin-flip
algorithms (TW1, TW0) underwent a similar degree of effort in optimisation. They work in
an analogous way as far as it makes sense for them to do so. TW0 is optimised further in
the following way, eliminating all arithmetic operations for a given $\beta$, it only
requires a simple lookup of the neighbours of a spin to get the probability that the new
spin should be up or down, whereas for TW1 it appears to be actually necessary to
accumulate Z-values. The arithmetic involved in TW1 can, however, be reduced to a few
simple multiplications, additions and divisions, with no exponentials or logarithms
necessary in the inner loops, due to a fortunate way in which the required numerical range
is locally bounded: see Appendix~\ref{AppendixB} for further details. The level of
optimisation used here is not as great as with the fastest multispin implementations
described in \cite{isakov}, though the code is general enough to work just as well with
arbitrary weights $J_{ij}$. In the language of \cite{isakov}, the TW0 code used here
achieves about $0.16$ spin-flips per nanosecond using a single thread.

The timings for the implementation and computer used are given in Appendix~\ref{AppendixA}.

\subsection{Results of equilibration comparison}

\begin{table}[h]
\begin{center}
\begin{tabular}{r r r r r r r r r}
Chimera size & $N$ & $n_\text{TW0}$ & $t^\text{eq}_\text{TW0}/\text{s}$ &
$n_\text{TW1}$ & $t^\text{eq}_\text{TW1}/\text{s}$ &
$t^\text{eq}_\text{TW0}/t^\text{eq}_\text{TW1}$ \\
\hline
$6\times6$ & $288$ & $3.76\times10^3$ & $0.0549$ & $68.5$ & $0.0126$ & $4.36$ \\
$8\times8$ & $512$ & $2.24\times10^4$ & $0.804$ & $303$ & $0.136$ & $5.91$ \\
$10\times10$ & $800$ & $1.49\times10^5$ & $10.7$ & $1.24\times10^3$ & $1.26$ & $8.44$ \\
$12\times12$ & $1152$ & $6.00\times10^5$ & $85.8$ & $3.25\times10^3$ & $6.05$ & $14.2$
\end{tabular}
\end{center}
\caption{Results. $n_\text{TW0}$ and $n_\text{TW1}$ denote the number of equilibration steps required, and the last column gives the time advantage of TW1 over TW0.}
\label{t2}
\end{table}

Table~\ref{t2} shows a modest but potentially useful speed improvement (last column) which
appears to increase with problem size. The factor of 14 is small enough that it could
potentially be erased by a better implementation of TW0, but on the face of it it is worth
having, and may increase for larger problem sizes.

\section{Comparison - finding ground states using parallel tempering}\label{comparison3}

Four methods are compared here, named GS-TW1, GS-TW2, PT-TW0 and PT-TW1. GS-TW1 and GS-TW2
are specialised ground state finding algorithms described in Appendix B of
\cite{selbydwave1} (where GS-TW1 corresponds to strategies 3 and 13, and GS-TW2
corresponds to strategies 4 and 14) and reviewed in subsection \ref{gsmethod} below, using
treewidth 1 and 2 subgraph neighbourhoods respectively as measured in big vertices,
roughly equivalent to treewidth 4 and 8 in terms of single spins (though slightly more
powerful than that, due to handling half of the $K_{4,4}$ as a single unit).

PT-TW0 and PT-TW1 are parallel tempering each with a fixed set of temperatures (for each
graph size), where the Monte Carlo move updates, respectively, the spin variables of a
subgraph of treewidth 0 (conventional update based on immediate neighbours) or one of
treewidth 1 (tree-based Monte Carlo move as described in Section~\ref{description}). The
treewidths 0 and 1 are in terms of big vertices and are roughly equivalent to treewidths 0
and 4 in terms of single spins (though, as before, slightly more powerful than that).

\subsection{Ground state finding method GS-TWw ($w=0,1,2,\ldots$)}\label{gsmethod}

This is the basic method we use to search for low energy states or ground states, subject
to adjustments as noted below. $E(S)$ denotes the energy of state $S$.  As noted above,
treewidth $w$ refers to big vertices, so is approximately equivalent to treewidth $4w$ on
individual spins.

\begin{enumerate}
\item Set $A=\{\}$, a multiset, and randomise the current state (configuration of spin variables), $\SC$.
\item Randomly perturb $\SC$ and let $B=0$.
\item Let $E_0=E(\SC)$. Repeatedly loop through each subgraph in the treewidth $w$
  subgraph collection and change $\SC$ by performing subgraph updates at
  $\beta=\infty$ until you've done a sweep that doesn't lower the energy. Let
  $E=E(\SC)$.
\item If $E<E_0$ then let $A=A\cup\{E\}$, $B=0$.
\item Increase $B$ by $|\{a\in A|a\le E\}|$. If $B/|A|$ is
below a certain threshold then go back to step 3, otherwise go to step 2.
\end{enumerate}

It is worth remembering that even at $\beta=\infty$ (finding the minimum energy of the
subgraph) the subgraph update is random because it has to make choices between equal
energy substates as it runs.

The subgraph collections used are fixed and chosen by hand, as illustrated in
Appendix~\ref{AppendixC}.

The idea of the $B$ counter is that $|\{a\in A|a\le E\}|/|A|$ approximates the
probability, $p$, that a random state being considered has lower energy than the current
one, and the above method tries to spend time proportional to $1/p$ considering states
which are $p$ of the way up the energy distribution. In that way, it tries to spend a
longer time working on the more promising states.

This is an idealised non-terminating version of the algorithm, finding lower and lower
energies indefinitely.  The time-to-solve (TTS) of GS-TWw is defined by the average time
it takes this version to find the ground state energy, as if there is an oracle that knows
the answer and can halt the solver when it has found it. The solver is not required to
state any particular level of confidence in its answer, only to arrive at it.

In practice, the above algorithm is modified to terminate at or below some target energy
$E_t$. It is then called with successively lower values of $E_t$, each time returning the
lowest energy found. The timing evaluator then looks something like this:

\begin{enumerate}
\item Let $E_t=\infty$
\item Let $n=0$, $\text{timer}=0$
\item Repeat until $n=500$:
\item \hskip3em Call above algorithm, modified to terminate at or below energy $E_t$. Let $E$ be the minimum energy obtained during this run.
\item \hskip3em If $E<E_t$, let $E_t=E$ and goto step 2
\item \hskip3em Let $n=n+1$
\item Result is $n$ instances of energy $E_t$ found in the time interval $T$ since $\text{timer}$ was last set to 0, and the estimated TTS is $T/n$.
\end{enumerate}

The random number generator is never reset, so that runs should be independent. If the
final energy obtained is $E_{\min}$, then a state with energy $E_{\min}$ will have been
encountered $n+1$ times during the process. The first occurrence has to be discarded for
timing purposes since it was obtained before it was known that $E_{\min}$ was the minimum
energy, and so runs were being interrupted and restarted.

Waiting for $500$ minimum energy states serves two purposes. First, it makes the estimated
TTS more accurate, both statistically and also by overcoming the accuracy limitations of
the CPU timer on the computer for fast cases. Second, it allows one to be reasonably
confident that it has found the actual ground state rather than just a low energy
state. Experimentally it appears that the energy landscape is sufficiently well-behaved
that during the course of using the above algorithm to look for an energy $E>E_0$, where
$E_0$ is the ground state energy, there is a reasonable probability (at least $0.05$ or so
for Range 7, and usually much lower) that it will encounter an energy less than $E$, and
these probabilities are independent for different runs. This means that finding $500$
independent minimum energies should ensure that there is only a small chance
$0.95^{500}\approx7\times10^{-12}$ of having missed the true ground state. Of course this
is by no means a rigorous argument and there remains the possibility of hidden bad cases
arising with larger $n$.  However, this regularity hypothesis has been verified for
$n\le8$ where exact ground states can be found, and for $n>8$ it is at least still valid
to compare the various solvers by requiring each to find the same lowest energy state
found by any of the solvers.

\subsection{Results of ground state finding}

The comparison was made on Range 7 instances, which are harder and perhaps better for
comparison purposes than Range 1 instances as they don't have a large artificial
degeneracy.  GS-TW1 and GS-TW2 were run for $n=4,5,\ldots,16$. For $n<10$, 1000 instances
were used; for $n=10$ and $n=11$, 500 instances were used; for $n=12$ and $n=13$, 250
instances were used; and for $n>13$, 100 instances were used.  PT-TW0 and PT-TW1 were run
for even values of $n$ from $4$ to $16$ inclusive. For $n<10$, 1000 instances were used;
for $n=12$, 250 instances were used; and for $n>12$, 100 instances were used.

\begin{figure}[h]
\input{range7.tex}
\caption{Comparison of four methods of finding ground states}
\label{range7graph}
\end{figure}
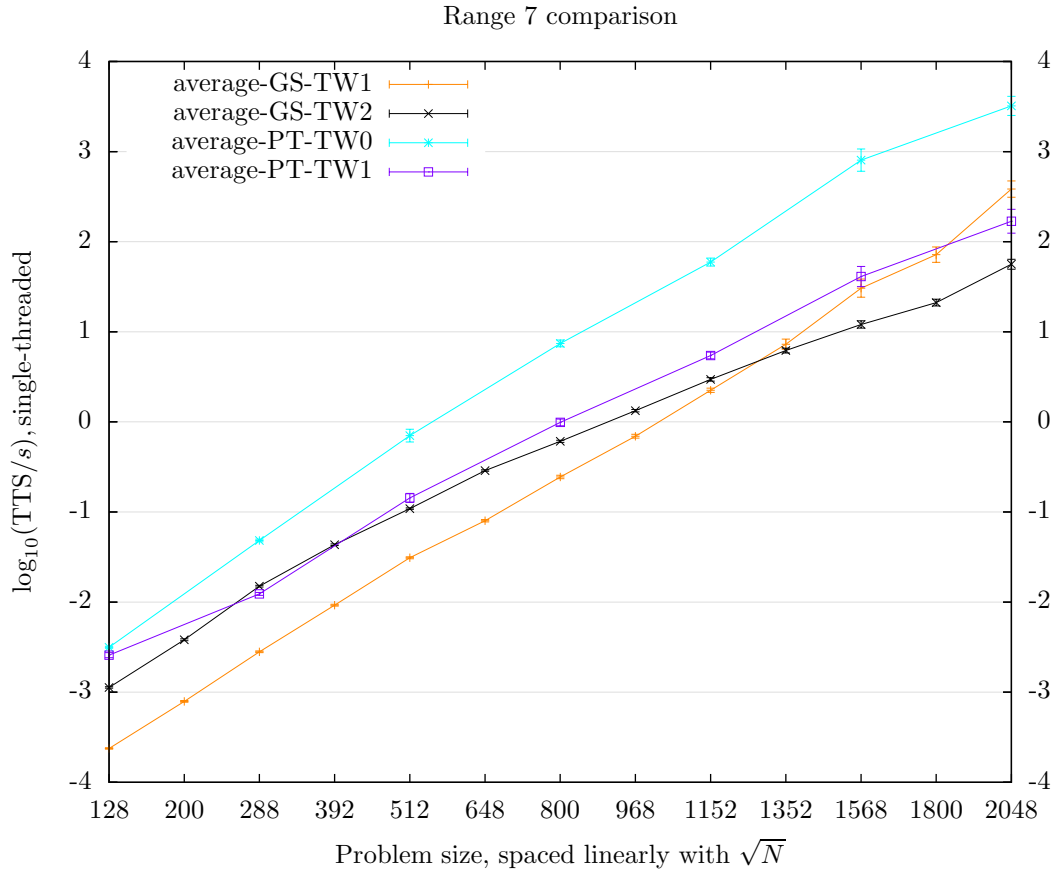

The results are shown in Fig.~\ref{range7graph}. The graphs of log-time vs linear size are
approximately linear (though perhaps slightly concave) with different gradients in each
case. PT-TW1 outperforms PT-TW0 by some margin, and the gap appears to widen with problem
size. Interestingly, PT-TW1 crosses over GS-TW1, showing the effectiveness of parallel
tempering as a meta-technique. GS-TW2 is initially worse than GS-TW1 for the easier
problems, but crosses over at approximately $n=13$ ($N=1352$), showing the value of using
moderately high treewidth (GS-TW2 has a treewidth of 8 in terms of individual spins).

\section{Conclusions and discussion}

The above results show some evidence for an advantage of using subgraph-based methods over
traditional spin-flip methods, and also some evidence that this advantage increases with
problem size. The question arises as to how representative these results are of more
general problems on more general graphs.

It is possible that a constant factor of this advantage could be erased by improvements to
the low-level SSS spin-flip, such as using a GPU, that might not be applicable to the SGS
case. On the other hand, if the problem were generalised slightly, for example by using
more than two spin states, then the extra complexity may hit SSS harder than SGS.

It is possible that a dynamically adaptive method of choosing temperatures, as mentioned
in \cite{katzgraberemc}, would help SSS more than SGS, because it might be especially
helpful with the difficult disorders for which SGS has a greater advantage over SSS. It is
hard to make a confident guess at the differential advantage, and so this needs to be
tested.

On the other hand, the Chimera graph is in fact a relatively easy graph to simulate
because it is somewhat sparse and locally-connected. Since the advantage of SGS over SSS
appears to be larger with the more difficult disorders and larger problem sizes, it is
possible that a more difficult graph altogether would show the advantage of SGS over
SSS considerably more strongly, though this can't be taken for granted as the subgraphs
used would also become more restricted. An interesting next experiment would be to see how
well SGS methods, using different treewidths, perform on a 3D spin glass.

It would be interesting to apply the methods described here to Markov Random Fields,
possibly giving more efficient inference in certain difficult cases. This was the
motivating factor in \cite{hamze} and \cite{fix}.

\begin{appendices}
\section{Further details of equilibration comparison}\label{AppendixA}

\begin{table}[h]
\begin{center}
\begin{tabular}{|p{7em}|p{10em}|p{14em}|}
\hline
Chimera size & $\beta$s used by SSS only & $\beta$s used by SSS and SGS\\
\hline
$6\times6$ & 0.256 0.296 0.346  & 0.414 0.507 0.645 0.882 1.375 2.598 20.000 \\
$8\times8$ & 0.285 0.322  & 0.363 0.414 0.478 0.559 0.669 0.840 1.121 1.646 2.894 20.000 \\
$10\times10$ & 0.310  & 0.341 0.376 0.419 0.472 0.539 0.622 0.736 0.892 1.134 1.531 2.222 3.507 6.166 20.000 \\
$12\times12$ & 0.250 0.269 0.289  & 0.310 0.337 0.367 0.399 0.439 0.489 0.552 0.630 0.727 0.850 1.018 1.264 1.626 2.169 3.110 4.850 20.000 \\
$14\times14$ & 0.244 0.259 0.275 0.292 0.310 & 0.333 0.358 0.385 0.419 0.455 0.495 0.545 0.608 0.677 0.763 0.871 1.006 1.190 1.442 1.812 2.417 3.724 20.000 \\
\hline
\end{tabular}
\end{center}
\caption{Temperature sets used}
\label{t3}
\end{table}

\begin{table}[h]
\begin{center}
\begin{tabular}{r r r r r r r}
Chimera size & $N$ & $t_\text{SSS}/\mu\text{s}$ & $nt_\text{SSS}$ & $t_\text{SGS}/\mu\text{s}$ & $nt_\text{SGS}$ & $t_\text{SGS}/t_\text{SSS}$\\
\hline
$6\times6$ & $288$ & $14.6$ & $10$ & $184$ & $7$ & $12.6$ \\
$8\times8$ & $512$ & $35.9$ & $12$ & $449$ & $10$ & $12.5$ \\
$10\times10$ & $800$ & $71.7$ & $15$ & $1020$ & $14$ & $14.2$ \\
$12\times12$ & $1152$ & $143$ & $20$ & $1860$ & $17$ & $13.0$ \\
$14\times14$ & $1568$ & $229$ & $23$ & $2800$ & $18$ & $12.2$ \\
\end{tabular}
\end{center}
\caption{Setup and low-level reference timings. $nt_X$ is the number of temperatures used for method $X$.}
\label{t1}
\end{table}

\section{Numerical considerations}\label{AppendixB}

Given the basic method of Section~\ref{description} to compute the $Z$-values of the
partial subgraphs, there are still a number of ways of implementing it. If the values of
$\log(Z)$ are stored instead of $Z$ then it will be necessary to perform logarithms and
exponentials in the inner loop, which will be slow. (It would not be possible, in all but
the simplest cases, to use look-up tables to eliminate this, since the number of entries
in such a table would be exponential in the number of neighbouring vertices of the
subgraph.)

Instead of storing $\log(Z)$, the implementation here stores the $Z$-values themselves.
This means that it is possible to avoid all logarithms and exponentials in the inner loop
by using a look-up table of $e^{-\beta n}$ for the relatively manageable number of
different $\beta$s and $n$ arising from energy differences from a single vertex. (A slight
enhancement is to use look-up tables to precalculate values of $n$ arising from a small
neighbourhood of a vertex, rather than from the single vertex itself.) Using $Z$-values in
this way is quite convenient and means that the inner loop calculation involves only a few
lookups, multiplications and additions. However, there is a potential problem that the
values of $Z$ can become too large or small for a particular floating point
representation. For speed, one would ideally like to be able to use a native
representation, typically IEE 754 binary64 (double) or 80-bit extended precision format
native to the x86 processor family.

Recall from Section~\ref{description}, sampling a spin configuration on a subgraph $T$
proceeds by constructing intermediate partition functions, $Z_H(\SC_{\partial H})$, of a
series of subgraphs $H\subset T$. The partition function $Z_H(\SC_{\partial H})$ depends
on the spin configuration on $\partial H$, where $\partial H$ is the ``boundary within
$T$'', the set of vertices of $T$ adjacent to $H$ that are not in $H$ (and also on
$\SC_{G\setminus T}$, though this latter dependency is suppressed as $\SC_{G\setminus T}$
is constant during the sampling process associated with $T$). The question arises: how
much precision (size of mantissa) and range (size of exponent) is required to store
$Z_H(\SC_{\partial H})$ adequately?

The precision presents no problem because all $Z$-values are \emph{positive}-linear
combinations of other $Z$-values. At each stage of the algorithm a random choice of spins
will be made based on the relative values of $Z_H(\SC_{\partial H})$, that is the spin
configuration $\SC_{\partial H}$ is chosen with probability $Z_H(\SC_{\partial
  H})/\sum_{\SC'_{\partial H}}Z_H(\SC'_{\partial H})$.  Thus to make an accurate random
choice, it is sufficient to know $Z_H(\SC_{\partial H})$ to a modest relative accuracy of,
say, $10^{-9}$. Because these $Z$-values are positive-linear combinations of other
$Z$-values from other subgraphs, it is sufficient to know these other $Z$-values to a
relative accuracy of $10^{-9}$.

Turning to the required size of the exponent of the floating point representation, we
first observe that the overall normalisation of the $Z$-values doesn't matter. That is,
$Z_H(\SC_{\partial H})$ can be multiplied by an overall factor $\lambda_H$, depending on
$H$ but not on $\SC_{\partial H}$. This means that we are not worried if all of the
$Z$-values simultaneously become very large or small as $H$ grows. However, on the face of
it, it is possible that for a given $H$ the relative values for different spin
configurations, $Z_H(\SC_{\partial H})/Z_H(\SC'_{\partial H})$, might be too big or small,
which would present a problem no matter how the $Z$-values were rescaled.

Fortunately at this point we are rescued by a convenient fact. The range of $Z$-values for
a given $H$ is limited by construction to a locally-computable constant and does not grow
with the size of the partial subgraph it is defined on.

To see this, let us first define $E(\SC_H)$ to be the total energy of the spin
configuration $\SC_H$ taken along edges within $H$ and along edges between $H$ and
$G\setminus T$, and $E(\SC_{\partial H};\SC_H)$ to be the total energy of the spin
configurations $\SC_{\partial H}$ and $\SC_H$ taken along edges joining $\partial H$ to
$H\cup \partial H\cup (G\setminus T)$.  Then
\[Z_H(\SC_{\partial H})=\sum_{\SC_H}e^{-\beta (E(\SC_H)+E(\SC_{\partial H};\SC_H))}.
\]
If $M(H)$ is defined by
\[
M(H)=\max_{\SC_H, \SC_{\partial H}, \SC'_{\partial H}}|E(\SC_{\partial H};\SC_H)-E(\SC'_{\partial H};\SC_H))|,
\]
then for two different boundary spin configurations $\SC_{\partial H}$ and
$\SC'_{\partial H}$, we have
\[
e^{-\beta M(H)}\le Z(\SC'_{\partial H})/Z(\SC_{\partial H})\le e^{\beta M(H)}.
\]
Note also that $E(\SC_{\partial H};\SC_H)=E(\SC_{\partial H};\SC_{H\cap\partial \partial
  H})$, so $E(\SC_{\partial H};\SC_H)$ does not notice what goes on deep within $H$.  The
choice of subgraphs will typically be designed to minimise the boundary $\partial H$, and
for the examples of $T$ in this preprint, $M(H)$ is defined by a local calculation
involving only a few vertices.

The maximum $\beta$ that can safely be used with this simple floating point representation
of $Z$ is determined by the maximum value of $M(H)$ over subgraphs $H$ of $T$ that are
used in the tree decomposition, the maximum degree of a node in the tree decomposition and
the size of the exponent in the floating point representation. The actual calculations of
$M(H)$ are quite tedious and omitted here, but in the examples used in this preprint it
was found that by using extended precision floating point representation, $\beta^*$ (the
largest inverse temperature that it was necessary to use; see Section~\ref{ptparams}) was
comfortably smaller than the maximum value of $\beta$ not causing calculations to
overflow.

\section{Subgraphs used for GS-TW1 and GS-TW2}\label{AppendixC}

\begin{figure}[h]
\centering
\includegraphics{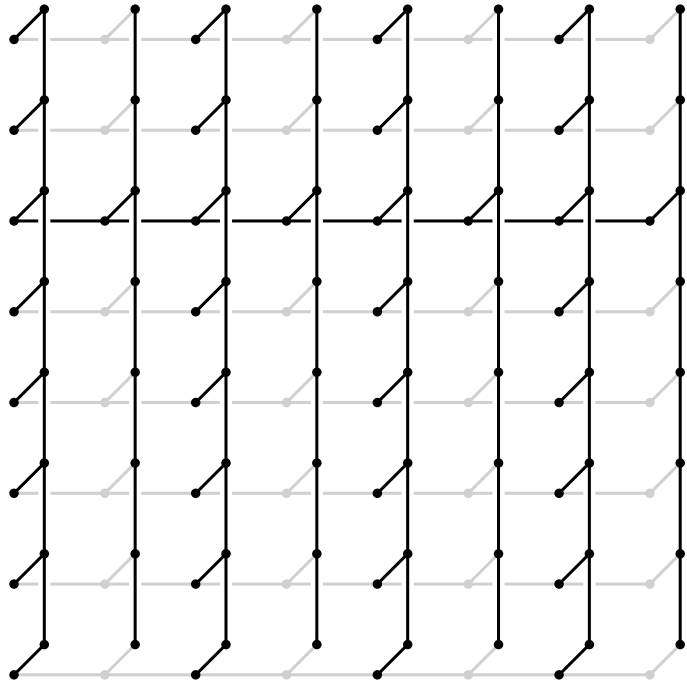}
\caption{A maximal induced tree (shown in dark) on the ``big vertex'' graph used for
  method GS-TW1. Including the versions of this subgraph with horizontals exchanged for
  verticals, there are 32 possible sets of this form. Coverage = 100/128 = 78.1\% of
  ``big vertices''.}
\label{C8S3}
\end{figure}

\begin{figure}[h]
\centering
\includegraphics{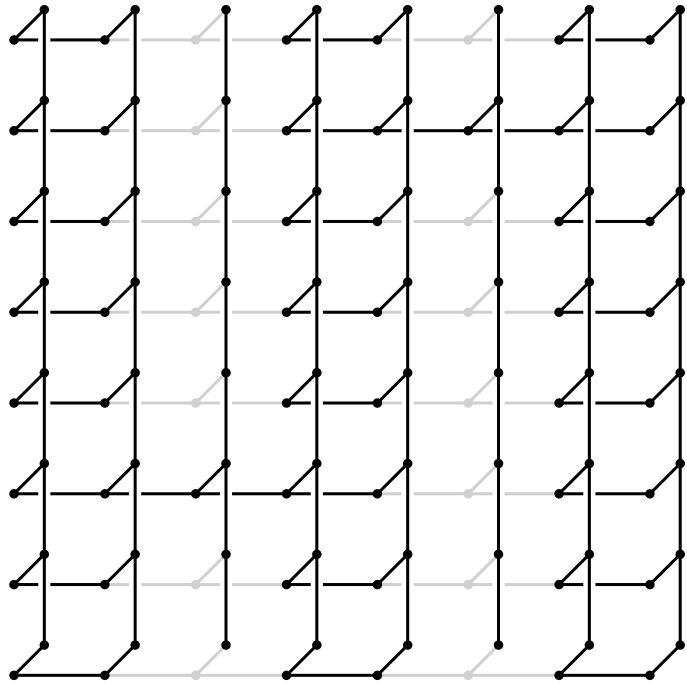}
\caption{A maximal induced treewidth-2 subgraph used for method GS-TW2. Coverage = 114/128 = 89.1\% of ``big vertices''.}
\label{C8S4}
\end{figure}

\begin{figure}[h]
\centering
\includegraphics{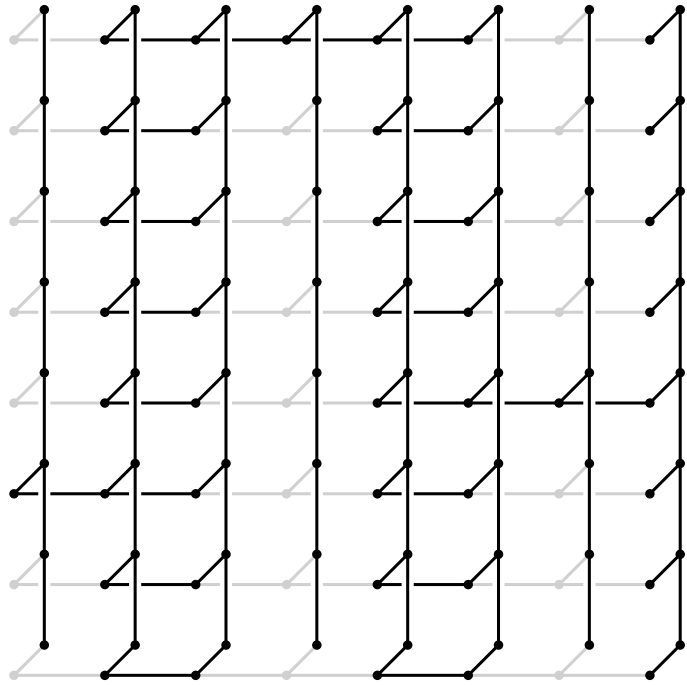}
\caption{Another maximal induced treewidth-2 subgraph used for method GS-TW2. Coverage = 107/128 = 83.6\% of ``big vertices''.}
\label{C8S4p1}
\end{figure}

\end{appendices}

\end{document}

%% file: Comparison-Katzgraber.tex
\begingroup
  \makeatletter
  \providecommand\color[2][]{%
    \GenericError{(gnuplot) \space\space\space\@spaces}{%
      Package color not loaded in conjunction with
      terminal option `colourtext'%
    }{See the gnuplot documentation for explanation.%
    }{Either use 'blacktext' in gnuplot or load the package
      color.sty in LaTeX.}%
    \renewcommand\color[2][]{}%
  }%
  \providecommand\includegraphics[2][]{%
    \GenericError{(gnuplot) \space\space\space\@spaces}{%
      Package graphicx or graphics not loaded%
    }{See the gnuplot documentation for explanation.%
    }{The gnuplot epslatex terminal needs graphicx.sty or graphics.sty.}%
    \renewcommand\includegraphics[2][]{}%
  }%
  \providecommand\rotatebox[2]{#2}%
  \@ifundefined{ifGPcolor}{%
    \newif\ifGPcolor
    \GPcolortrue
  }{}%
  \@ifundefined{ifGPblacktext}{%
    \newif\ifGPblacktext
    \GPblacktexttrue
  }{}%
  \let\gplgaddtomacro\g@addto@macro
  \gdef\gplbacktext{}%
  \gdef\gplfronttext{}%
  \makeatother
  \ifGPblacktext
    \def\colorrgb#1{}%
    \def\colorgray#1{}%
  \else
    \ifGPcolor
      \def\colorrgb#1{\color[rgb]{#1}}%
      \def\colorgray#1{\color[gray]{#1}}%
      \expandafter\def\csname LTw\endcsname{\color{white}}%
      \expandafter\def\csname LTb\endcsname{\color{black}}%
      \expandafter\def\csname LTa\endcsname{\color{black}}%
      \expandafter\def\csname LT0\endcsname{\color[rgb]{1,0,0}}%
      \expandafter\def\csname LT1\endcsname{\color[rgb]{0,1,0}}%
      \expandafter\def\csname LT2\endcsname{\color[rgb]{0,0,1}}%
      \expandafter\def\csname LT3\endcsname{\color[rgb]{1,0,1}}%
      \expandafter\def\csname LT4\endcsname{\color[rgb]{0,1,1}}%
      \expandafter\def\csname LT5\endcsname{\color[rgb]{1,1,0}}%
      \expandafter\def\csname LT6\endcsname{\color[rgb]{0,0,0}}%
      \expandafter\def\csname LT7\endcsname{\color[rgb]{1,0.3,0}}%
      \expandafter\def\csname LT8\endcsname{\color[rgb]{0.5,0.5,0.5}}%
    \else
      \def\colorrgb#1{\color{black}}%
      \def\colorgray#1{\color[gray]{#1}}%
      \expandafter\def\csname LTw\endcsname{\color{white}}%
      \expandafter\def\csname LTb\endcsname{\color{black}}%
      \expandafter\def\csname LTa\endcsname{\color{black}}%
      \expandafter\def\csname LT0\endcsname{\color{black}}%
      \expandafter\def\csname LT1\endcsname{\color{black}}%
      \expandafter\def\csname LT2\endcsname{\color{black}}%
      \expandafter\def\csname LT3\endcsname{\color{black}}%
      \expandafter\def\csname LT4\endcsname{\color{black}}%
      \expandafter\def\csname LT5\endcsname{\color{black}}%
      \expandafter\def\csname LT6\endcsname{\color{black}}%
      \expandafter\def\csname LT7\endcsname{\color{black}}%
      \expandafter\def\csname LT8\endcsname{\color{black}}%
    \fi
  \fi
  \setlength{\unitlength}{0.0500bp}%
  \begin{picture}(6236.00,4534.00)%
    \gplgaddtomacro\gplbacktext{%
      \csname LTb\endcsname%
      \put(-132,704){\makebox(0,0)[r]{\strut{}-4}}%
      \put(-132,1150){\makebox(0,0)[r]{\strut{}-3.5}}%
      \put(-132,1595){\makebox(0,0)[r]{\strut{}-3}}%
      \put(-132,2041){\makebox(0,0)[r]{\strut{}-2.5}}%
      \put(-132,2487){\makebox(0,0)[r]{\strut{}-2}}%
      \put(-132,2932){\makebox(0,0)[r]{\strut{}-1.5}}%
      \put(-132,3378){\makebox(0,0)[r]{\strut{}-1}}%
      \put(-132,3823){\makebox(0,0)[r]{\strut{}-0.5}}%
      \put(-132,4269){\makebox(0,0)[r]{\strut{} 0}}%
      \put(0,484){\makebox(0,0){\strut{} 0}}%
      \put(779,484){\makebox(0,0){\strut{} 0.5}}%
      \put(1559,484){\makebox(0,0){\strut{} 1}}%
      \put(2338,484){\makebox(0,0){\strut{} 1.5}}%
      \put(3118,484){\makebox(0,0){\strut{} 2}}%
      \put(3897,484){\makebox(0,0){\strut{} 2.5}}%
      \put(4676,484){\makebox(0,0){\strut{} 3}}%
      \put(5456,484){\makebox(0,0){\strut{} 3.5}}%
      \put(6235,484){\makebox(0,0){\strut{} 4}}%
      \put(-902,2486){\rotatebox{-270}{\makebox(0,0){\strut{}$\log_2g_q(T)$}}}%
      \put(3117,154){\makebox(0,0){\strut{}$T\cdot512^{1/8}$}}%
    }%
    \gplgaddtomacro\gplfronttext{%
      \csname LTb\endcsname%
      \put(5248,4096){\makebox(0,0)[r]{\strut{}\cite{katzgraber}, Fig. 2 bimodal}}%
      \csname LTb\endcsname%
      \put(5248,3876){\makebox(0,0)[r]{\strut{}Subgraph-based}}%
    }%
    \gplbacktext
    \put(0,0){\includegraphics{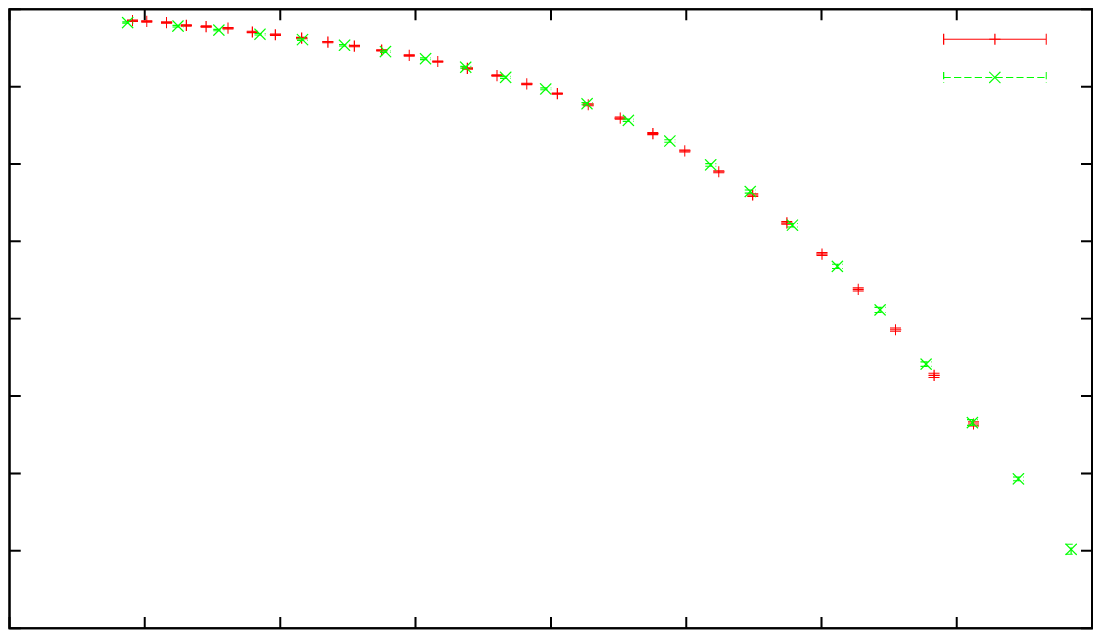}}%
    \gplfronttext
  \end{picture}%
\endgroup

%% file: Comparison-Katzgraber-errorsx10.tex
\begingroup
  \makeatletter
  \providecommand\color[2][]{%
    \GenericError{(gnuplot) \space\space\space\@spaces}{%
      Package color not loaded in conjunction with
      terminal option `colourtext'%
    }{See the gnuplot documentation for explanation.%
    }{Either use 'blacktext' in gnuplot or load the package
      color.sty in LaTeX.}%
    \renewcommand\color[2][]{}%
  }%
  \providecommand\includegraphics[2][]{%
    \GenericError{(gnuplot) \space\space\space\@spaces}{%
      Package graphicx or graphics not loaded%
    }{See the gnuplot documentation for explanation.%
    }{The gnuplot epslatex terminal needs graphicx.sty or graphics.sty.}%
    \renewcommand\includegraphics[2][]{}%
  }%
  \providecommand\rotatebox[2]{#2}%
  \@ifundefined{ifGPcolor}{%
    \newif\ifGPcolor
    \GPcolortrue
  }{}%
  \@ifundefined{ifGPblacktext}{%
    \newif\ifGPblacktext
    \GPblacktexttrue
  }{}%
  \let\gplgaddtomacro\g@addto@macro
  \gdef\gplbacktext{}%
  \gdef\gplfronttext{}%
  \makeatother
  \ifGPblacktext
    \def\colorrgb#1{}%
    \def\colorgray#1{}%
  \else
    \ifGPcolor
      \def\colorrgb#1{\color[rgb]{#1}}%
      \def\colorgray#1{\color[gray]{#1}}%
      \expandafter\def\csname LTw\endcsname{\color{white}}%
      \expandafter\def\csname LTb\endcsname{\color{black}}%
      \expandafter\def\csname LTa\endcsname{\color{black}}%
      \expandafter\def\csname LT0\endcsname{\color[rgb]{1,0,0}}%
      \expandafter\def\csname LT1\endcsname{\color[rgb]{0,1,0}}%
      \expandafter\def\csname LT2\endcsname{\color[rgb]{0,0,1}}%
      \expandafter\def\csname LT3\endcsname{\color[rgb]{1,0,1}}%
      \expandafter\def\csname LT4\endcsname{\color[rgb]{0,1,1}}%
      \expandafter\def\csname LT5\endcsname{\color[rgb]{1,1,0}}%
      \expandafter\def\csname LT6\endcsname{\color[rgb]{0,0,0}}%
      \expandafter\def\csname LT7\endcsname{\color[rgb]{1,0.3,0}}%
      \expandafter\def\csname LT8\endcsname{\color[rgb]{0.5,0.5,0.5}}%
    \else
      \def\colorrgb#1{\color{black}}%
      \def\colorgray#1{\color[gray]{#1}}%
      \expandafter\def\csname LTw\endcsname{\color{white}}%
      \expandafter\def\csname LTb\endcsname{\color{black}}%
      \expandafter\def\csname LTa\endcsname{\color{black}}%
      \expandafter\def\csname LT0\endcsname{\color{black}}%
      \expandafter\def\csname LT1\endcsname{\color{black}}%
      \expandafter\def\csname LT2\endcsname{\color{black}}%
      \expandafter\def\csname LT3\endcsname{\color{black}}%
      \expandafter\def\csname LT4\endcsname{\color{black}}%
      \expandafter\def\csname LT5\endcsname{\color{black}}%
      \expandafter\def\csname LT6\endcsname{\color{black}}%
      \expandafter\def\csname LT7\endcsname{\color{black}}%
      \expandafter\def\csname LT8\endcsname{\color{black}}%
    \fi
  \fi
  \setlength{\unitlength}{0.0500bp}%
  \begin{picture}(6236.00,4534.00)%
    \gplgaddtomacro\gplbacktext{%
      \csname LTb\endcsname%
      \put(-132,704){\makebox(0,0)[r]{\strut{}-4}}%
      \put(-132,1150){\makebox(0,0)[r]{\strut{}-3.5}}%
      \put(-132,1595){\makebox(0,0)[r]{\strut{}-3}}%
      \put(-132,2041){\makebox(0,0)[r]{\strut{}-2.5}}%
      \put(-132,2487){\makebox(0,0)[r]{\strut{}-2}}%
      \put(-132,2932){\makebox(0,0)[r]{\strut{}-1.5}}%
      \put(-132,3378){\makebox(0,0)[r]{\strut{}-1}}%
      \put(-132,3823){\makebox(0,0)[r]{\strut{}-0.5}}%
      \put(-132,4269){\makebox(0,0)[r]{\strut{} 0}}%
      \put(0,484){\makebox(0,0){\strut{} 0}}%
      \put(779,484){\makebox(0,0){\strut{} 0.5}}%
      \put(1559,484){\makebox(0,0){\strut{} 1}}%
      \put(2338,484){\makebox(0,0){\strut{} 1.5}}%
      \put(3118,484){\makebox(0,0){\strut{} 2}}%
      \put(3897,484){\makebox(0,0){\strut{} 2.5}}%
      \put(4676,484){\makebox(0,0){\strut{} 3}}%
      \put(5456,484){\makebox(0,0){\strut{} 3.5}}%
      \put(6235,484){\makebox(0,0){\strut{} 4}}%
      \put(-902,2486){\rotatebox{-270}{\makebox(0,0){\strut{}$\log_2g_q(T)$}}}%
      \put(3117,154){\makebox(0,0){\strut{}$T\cdot512^{1/8}$}}%
    }%
    \gplgaddtomacro\gplfronttext{%
      \csname LTb\endcsname%
      \put(5248,4096){\makebox(0,0)[r]{\strut{}\cite{katzgraber}, Fig. 2 bimodal}}%
      \csname LTb\endcsname%
      \put(5248,3876){\makebox(0,0)[r]{\strut{}Subgraph-based}}%
    }%
    \gplbacktext
    \put(0,0){\includegraphics{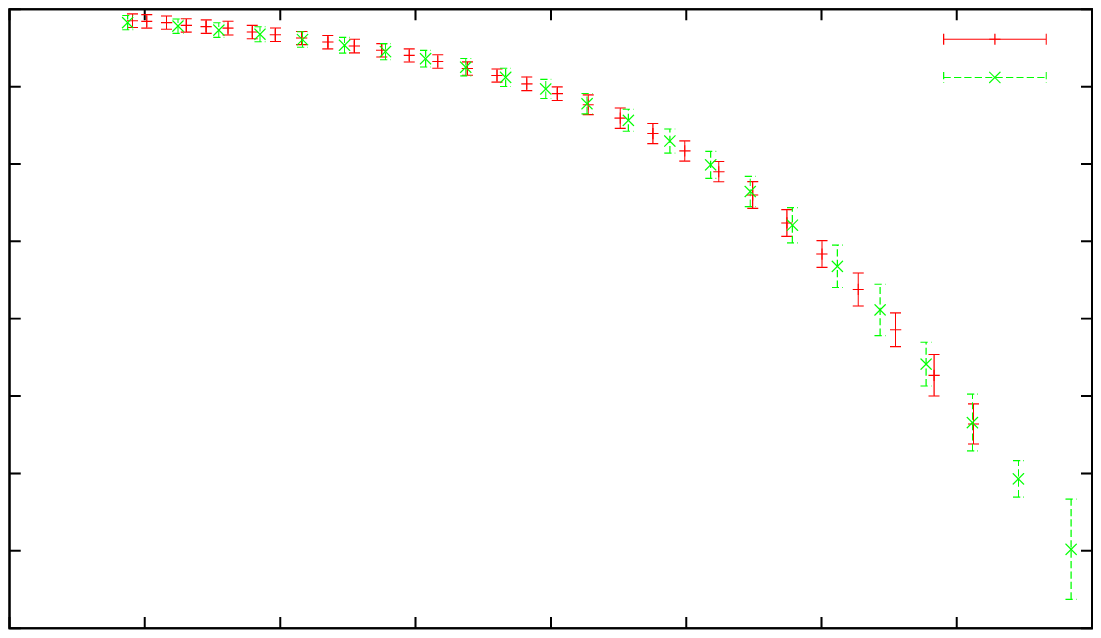}}%
    \gplfronttext
  \end{picture}%
\endgroup

%% file: range7.tex
\begingroup
  \makeatletter
  \providecommand\color[2][]{%
    \GenericError{(gnuplot) \space\space\space\@spaces}{%
      Package color not loaded in conjunction with
      terminal option `colourtext'%
    }{See the gnuplot documentation for explanation.%
    }{Either use 'blacktext' in gnuplot or load the package
      color.sty in LaTeX.}%
    \renewcommand\color[2][]{}%
  }%
  \providecommand\includegraphics[2][]{%
    \GenericError{(gnuplot) \space\space\space\@spaces}{%
      Package graphicx or graphics not loaded%
    }{See the gnuplot documentation for explanation.%
    }{The gnuplot epslatex terminal needs graphicx.sty or graphics.sty.}%
    \renewcommand\includegraphics[2][]{}%
  }%
  \providecommand\rotatebox[2]{#2}%
  \@ifundefined{ifGPcolor}{%
    \newif\ifGPcolor
    \GPcolortrue
  }{}%
  \@ifundefined{ifGPblacktext}{%
    \newif\ifGPblacktext
    \GPblacktexttrue
  }{}%
  \let\gplgaddtomacro\g@addto@macro
  \gdef\gplbacktext{}%
  \gdef\gplfronttext{}%
  \makeatother
  \ifGPblacktext
    \def\colorrgb#1{}%
    \def\colorgray#1{}%
  \else
    \ifGPcolor
      \def\colorrgb#1{\color[rgb]{#1}}%
      \def\colorgray#1{\color[gray]{#1}}%
      \expandafter\def\csname LTw\endcsname{\color{white}}%
      \expandafter\def\csname LTb\endcsname{\color{black}}%
      \expandafter\def\csname LTa\endcsname{\color{black}}%
      \expandafter\def\csname LT0\endcsname{\color[rgb]{1,0,0}}%
      \expandafter\def\csname LT1\endcsname{\color[rgb]{0,1,0}}%
      \expandafter\def\csname LT2\endcsname{\color[rgb]{0,0,1}}%
      \expandafter\def\csname LT3\endcsname{\color[rgb]{1,0,1}}%
      \expandafter\def\csname LT4\endcsname{\color[rgb]{0,1,1}}%
      \expandafter\def\csname LT5\endcsname{\color[rgb]{1,1,0}}%
      \expandafter\def\csname LT6\endcsname{\color[rgb]{0,0,0}}%
      \expandafter\def\csname LT7\endcsname{\color[rgb]{1,0.3,0}}%
      \expandafter\def\csname LT8\endcsname{\color[rgb]{0.5,0.5,0.5}}%
    \else
      \def\colorrgb#1{\color{black}}%
      \def\colorgray#1{\color[gray]{#1}}%
      \expandafter\def\csname LTw\endcsname{\color{white}}%
      \expandafter\def\csname LTb\endcsname{\color{black}}%
      \expandafter\def\csname LTa\endcsname{\color{black}}%
      \expandafter\def\csname LT0\endcsname{\color{black}}%
      \expandafter\def\csname LT1\endcsname{\color{black}}%
      \expandafter\def\csname LT2\endcsname{\color{black}}%
      \expandafter\def\csname LT3\endcsname{\color{black}}%
      \expandafter\def\csname LT4\endcsname{\color{black}}%
      \expandafter\def\csname LT5\endcsname{\color{black}}%
      \expandafter\def\csname LT6\endcsname{\color{black}}%
      \expandafter\def\csname LT7\endcsname{\color{black}}%
      \expandafter\def\csname LT8\endcsname{\color{black}}%
    \fi
  \fi
  \setlength{\unitlength}{0.0500bp}%
  \begin{picture}(6802.00,6802.00)%
    \gplgaddtomacro\gplbacktext{%
      \csname LTb\endcsname%
      \put(-132,704){\makebox(0,0)[r]{\strut{}-4}}%
      \csname LTb\endcsname%
      \put(-132,1384){\makebox(0,0)[r]{\strut{}-3}}%
      \csname LTb\endcsname%
      \put(-132,2063){\makebox(0,0)[r]{\strut{}-2}}%
      \csname LTb\endcsname%
      \put(-132,2743){\makebox(0,0)[r]{\strut{}-1}}%
      \csname LTb\endcsname%
      \put(-132,3423){\makebox(0,0)[r]{\strut{} 0}}%
      \csname LTb\endcsname%
      \put(-132,4102){\makebox(0,0)[r]{\strut{} 1}}%
      \csname LTb\endcsname%
      \put(-132,4782){\makebox(0,0)[r]{\strut{} 2}}%
      \csname LTb\endcsname%
      \put(-132,5461){\makebox(0,0)[r]{\strut{} 3}}%
      \csname LTb\endcsname%
      \put(-132,6141){\makebox(0,0)[r]{\strut{} 4}}%
      \put(0,484){\makebox(0,0){\strut{}128}}%
      \put(567,484){\makebox(0,0){\strut{}200}}%
      \put(1133,484){\makebox(0,0){\strut{}288}}%
      \put(1700,484){\makebox(0,0){\strut{}392}}%
      \put(2267,484){\makebox(0,0){\strut{}512}}%
      \put(2834,484){\makebox(0,0){\strut{}648}}%
      \put(3400,484){\makebox(0,0){\strut{}800}}%
      \put(3967,484){\makebox(0,0){\strut{}968}}%
      \put(4534,484){\makebox(0,0){\strut{}1152}}%
      \put(5101,484){\makebox(0,0){\strut{}1352}}%
      \put(5668,484){\makebox(0,0){\strut{}1568}}%
      \put(6234,484){\makebox(0,0){\strut{}1800}}%
      \put(6801,484){\makebox(0,0){\strut{}2048}}%
      \put(6933,704){\makebox(0,0)[l]{\strut{}-4}}%
      \put(6933,1384){\makebox(0,0)[l]{\strut{}-3}}%
      \put(6933,2063){\makebox(0,0)[l]{\strut{}-2}}%
      \put(6933,2743){\makebox(0,0)[l]{\strut{}-1}}%
      \put(6933,3423){\makebox(0,0)[l]{\strut{} 0}}%
      \put(6933,4102){\makebox(0,0)[l]{\strut{} 1}}%
      \put(6933,4782){\makebox(0,0)[l]{\strut{} 2}}%
      \put(6933,5461){\makebox(0,0)[l]{\strut{} 3}}%
      \put(6933,6141){\makebox(0,0)[l]{\strut{} 4}}%
      \put(-638,3422){\rotatebox{-270}{\makebox(0,0){\strut{}$\log_{10}(\text{TTS}/s), \text{single-threaded}$}}}%
      \put(3400,154){\makebox(0,0){\strut{}Problem size, spaced linearly with $\sqrt{N}$}}%
      \put(3400,6471){\makebox(0,0){\strut{}Range 7 comparison}}%
    }%
    \gplgaddtomacro\gplfronttext{%
      \csname LTb\endcsname%
      \put(1980,5968){\makebox(0,0)[r]{\strut{}average-GS-TW1}}%
      \csname LTb\endcsname%
      \put(1980,5748){\makebox(0,0)[r]{\strut{}average-GS-TW2}}%
      \csname LTb\endcsname%
      \put(1980,5528){\makebox(0,0)[r]{\strut{}average-PT-TW0}}%
      \csname LTb\endcsname%
      \put(1980,5308){\makebox(0,0)[r]{\strut{}average-PT-TW1}}%
    }%
    \gplbacktext
    \put(0,0){\includegraphics{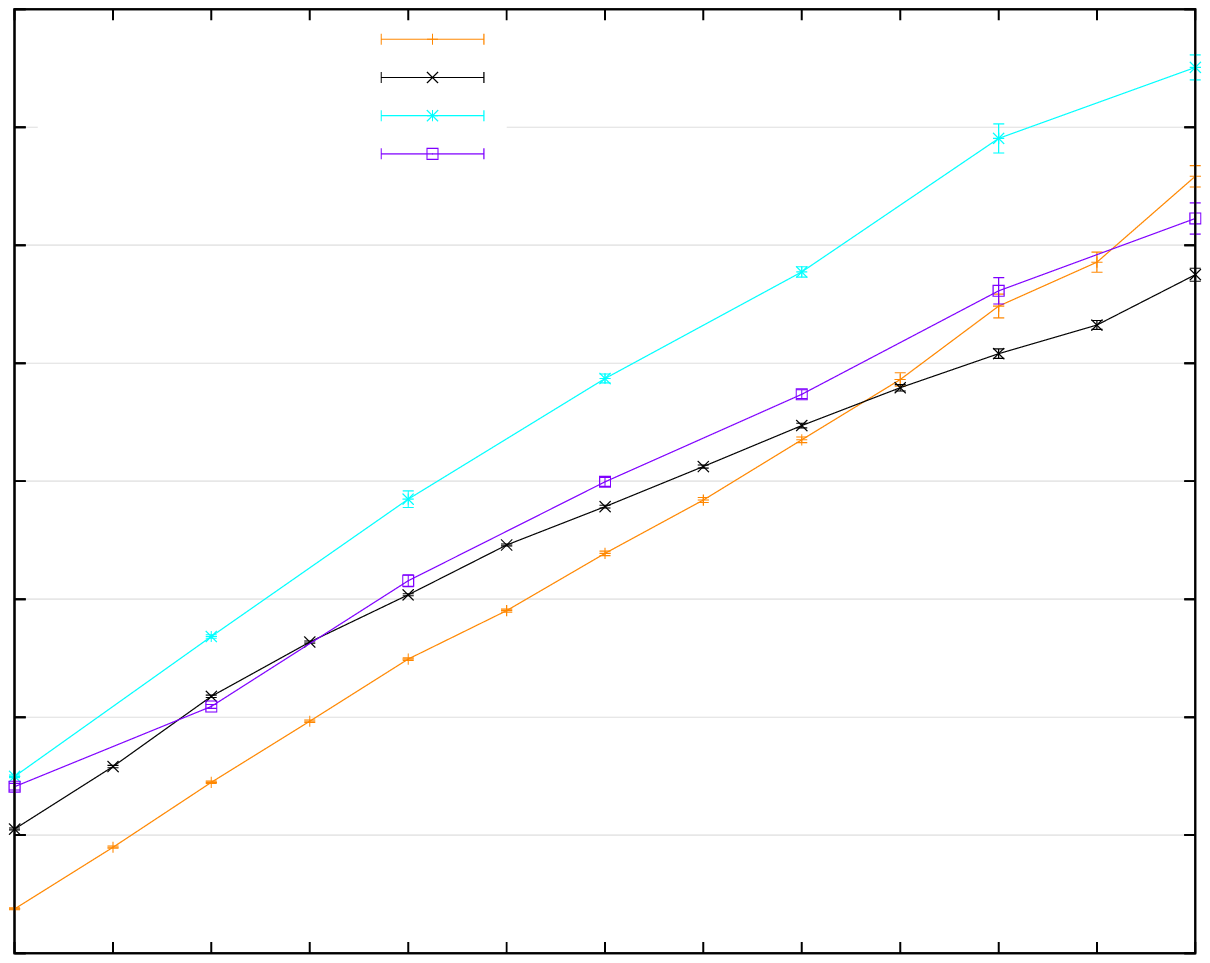}}%
    \gplfronttext
  \end{picture}%
\endgroup